# Lineshape-asymmetry elimination in weak atomic transitions driven by an intense standing wave field

Dionysios Antypas[1,*], Anne Fabricant[2], and Dmitry Budker[1,2,3]

[1]Helmholtz-Institut Mainz, Mainz 55128, Germany
[2]Johannes Gutenberg-Universität Mainz, Mainz 55128, Germany
[3]Department of Physics, University of California at Berkeley, Berkeley, CA 94720-7300, USA
*Corresponding author: dantypas@uni-mainz.de

**Owing to the ac-Stark effect, the lineshape of a weak optical transition in an atomic beam can become significantly distorted, when driven by an intense standing wave field. We use an Yb atomic beam to study the lineshape of the $6s^2\ ^1S_0 \rightarrow 5d6s\ ^3D_1$ transition, which is excited with light circulating in a Fabry-Perot resonator. We demonstrate two methods to avoid the distortion of the transition profile. Of these, one relies on the operation of the resonator in multiple longitudinal modes, and the other in multiple transverse modes.**

In spectroscopy of optical transitions with an atomic beam, the commonly observed spectral feature is a symmetric resonance profile. A rather peculiar situation can occur, however, when the transition is induced by an intense standing wave field. Owing to the imperfect collimation of an atomic beam, most of the atoms that travel nominally perpendicular to the light go through many intensity nodes and anti-nodes while traversing the standing wave. In the presence of an off-resonant ac-Stark effect, the moving atoms experience amplitude and effectively frequency modulation (due to the atomic energy-level modulation by the ac-Stark effect) that can result in a complex spectral profile for the resonance. This effect was first observed in the $6S_{1/2} \rightarrow 7S_{1/2}$ transition in Cs [1], and was thereafter studied extensively in $6s^2\ ^1S_0 \rightarrow 5d6s\ ^3D_1$ transition in Yb [2,3], due to its relevance to atomic parity violation (APV) studies in the same transition [4,5], and because the asymmetric lineshape can be analyzed to determine related ac-polarizabilities.

Since the occurring lineshape distortion is a potential source of systematic errors in our ongoing APV experiment [5], its elimination offers a simplification and reduction of such errors. In addition, since our APV experiment will perform measurements on a chain of Yb isotopes that have partially overlapping transitions, working with a symmetric profile further simplifies the analysis of data.

Here we demonstrate two methods to restore the symmetric profile of the $^{174}$Yb $^1S_0 \rightarrow\ ^3D_1$ transition, which for the typical optical power in our APV studies (≈30 W) is significantly distorted. The first relies on injection of several frequencies into the Fabry-Perot (FP) resonator which we employ as a power-build-up cavity (PBC). The second method involves the operation of the PBC in a configuration that allows simultaneous coupling of light to multiple transverse modes.

The operation of a FP in multiple longitudinal modes has been employed in a number of applications, including cavity-enhanced frequency-modulation spectroscopy [6,7], cavity-enhanced direct frequency comb spectroscopy [8,9], and broadband cavity ring-down spectroscopy [10,11]. A FP operation in multiple frequencies was used to demonstrate transformation of an optical dipole trap from a lattice configuration to that of an effectively traveling wave configuration [12], and more recently, a FP resonant with two frequencies was employed in the implementation of an intracavity atomic interferometer [13]. In relation to our lineshape-asymmetry studies, we follow a proposal by Bennett [14] and show that the injection of several frequencies into the PBC can create intracavity conditions under which this asymmetry nearly vanishes. To realize such conditions, we use an electro-optic modulator (EOM) to impose frequency sidebands on the light coupled to the fundamental transverse electromagnetic mode (TEM$_{00}$) of the PBC [15]. These sidebands are spaced from the carrier frequency $v_0$ by a free spectral range (FSR) of the cavity, such that in addition to $v_0$, 1st- and 2nd-order spectral components (of frequency $v=v_0\pm n\cdot$FSR, where $n$ is the sideband order) are also resonant in the PBC. Although only the carrier is resonant with the atoms, the presence of these additional frequencies can result in the elimination of spatial intensity variation at the PBC center. To illustrate this, we consider the intracavity intensity profile when frequency components $v=v_0\pm n\cdot$FSR (n=0,1,2) are present in the FP. The time-averaged intensity for the resulting standing wave (to within an overall amplitude factor) is given by [14]:

$$\langle I \rangle_t = \sin^2 kx[J_0^2(\delta) + 2J_1^2(\delta)\cos 2\beta x + 2J_2^2(\delta)\cos 4\beta x] + 2J_1^2(\delta)\sin^2\beta x + 2J_2^2(\delta)\sin^2 2\beta x, \quad (1)$$

where $J_n$ are Bessel functions of the first kind, $\delta$ is the modulation index that quantifies the relative amplitudes of the various frequency components, $k=2\pi/\lambda$ (where $\lambda$ is the carrier wavelength), $\beta=2\pi FSR/c$, and $c$ is the speed of light. (1) contains the rapidly oscillating term $sin^2kx$ (with period $\lambda/2$), the only term surviving in the absence of phase modulation ($\delta=0$), as well as several slow-varying terms with phases $2n\beta x$. By appropriate choice of $\delta$, an intensity profile is obtained for which the rapid spatial variations of $sin^2kx$ vanish at the FP center. When spectral components of sideband order n=0, 1, 2 circulate in the PBC, this value is expected to be $\delta=1.2$. We show in fig. 1 plots of the intensity envelope for $\delta=1.2$ as well as for $\delta$ values slightly lower or higher than 1.2.

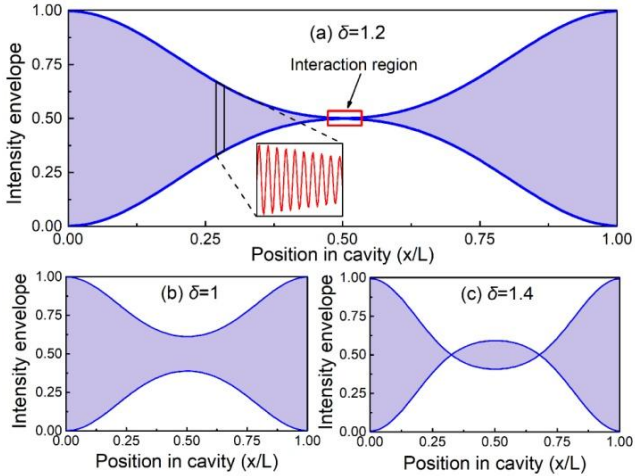

Fig. 1 (Color online). Simulated time-averaged intensity envelope plots for phase-modulated light injected into a PBC, for a modulation index a) $\delta=1.2$, b) $\delta=1$, and c) $\delta=1.4$. The gray areas indicate regions of rapid spatial variation of intensity.

A potential limitation to this method can be identified, considering the presence of other transitions in the vicinity of the resonance of interest. If no care is taken, a situation may arise where these adjacent transitions are excited by one of the sideband frequencies, resulting in unwanted contributions to the observed signal. This situation can be avoided by appropriate selection of the FP FSR, such that none of the sideband components overlaps with the frequency of a nearby transition.

Our second method to restore the symmetric profile of the $^1S_0 \rightarrow ^3D_1$ transition uses the PBC in a geometry similar to that known as the confocal FP [15], in which the spacing L between the cavity mirrors is made equal to their radius of curvature. The numerous TEM modes supported by a stable resonator of arbitrary length become degenerate for a confocal FP and the generally rich cavity spectrum reduces to a simple structure consisting of sets of two fringes. Each of these fringes is the result of large number of overlapping TEM modes, of either even or odd symmetry. The important aspect related to our lineshape studies is that the cavity beam waist in such a configuration can be made rather large, since input light can be coupled to a great number of TEM modes, thereby reducing the optical intensity and the associated effects of the ac-Stark shift on the lineshape.

The transverse mode degeneracy of the confocal FP can be satisfied for other geometries as well [16], in which the various TEM modes collapse into N groups in the spectrum (N=2 for the confocal FP), resulting in a FSR=c/2NL. Relative to a FP of arbitrary length, mode-matched to the TEM$_{00}$ mode, the power in each of the N fringes is N times lower.

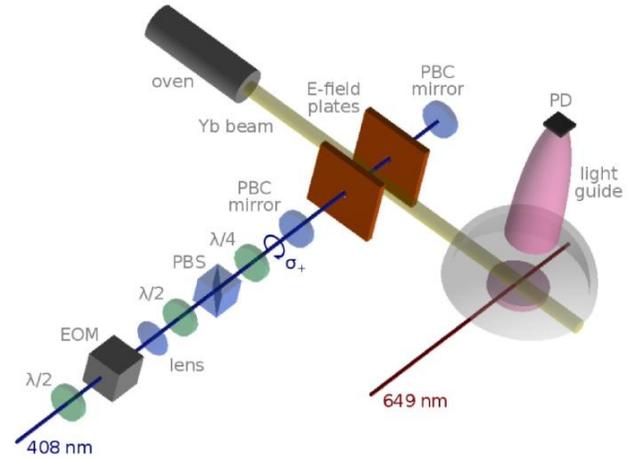

Fig. 2 (Color online). Schematic of the Yb atom-beam spectroscopy setup. $\lambda/2$: half-wave plate, $\lambda/4$: quarter-wave plate, PBS: polarizing beam- splitter, EOM: electro-optic modulator, PBC: power-build-up cavity, PD: photodiode.

The atomic-beam apparatus used in the spectroscopy of the Yb $^1S_0 \rightarrow ^3D_1$ transition is described in [5], and only certain details, mainly related to the PBC operation, are different here. We show a schematic of the setup in fig. 2. An Yb beam effuses from an oven heated to 530 °C. Atoms reach the interaction region with a mean longitudinal velocity of 277 m/s and a full width at half maximum (FWHM) transverse velocity spread of 8 m/s, where they intercept the standing wave of the PBC at normal incidence. As the $^1S_0 \rightarrow ^3D_1$ excitation involves states of same parity, the application of a dc electric field (2 kV/cm) to the atoms is necessary to allow an electric-dipole transition amplitude for the excitation [17]. No magnetic field is applied to the atoms; hence the different Zeeman components of the transition are unresolved. Most of the atoms undergoing this transition decay to the 6s6p $^3P_0$ metastable state, whose population is detected downstream from the interaction region. This is done through further excitation to the 6s7s $^3S_1$ state with $\approx$ 2 mW of laser light at 649 nm, and detection of the induced fluorescence with an optimized light-collector and a large-area photodiode [5]. The 649-nm light is produced with an external-cavity diode laser (Vitawave), whose frequency is stabilized to a wavemeter (High Finesse). The 408.3 nm light coupled to the PBC is produced with a frequency-doubled Ti:Sapphire laser system (M-Squared), outputting $\sim$1 W. Using frequency-modulation spectroscopy [18] and feedback to a piezo transducer mounted behind one of the PBC mirrors, the PBC length is stabilized to the peak of the cavity transmission. Transmission is also measured with a calibrated photodiode, in order to monitor the intracavity power. For the work presented here, up to 76 W of light was circulating in the PBC.

The PBC consists of a pair of mirrors with a 1 m radius of curvature and has a finesse of ≈500. In the experiment involving multiple-frequency injection into the PBC, the mirrors are spaced by L≈27.7 cm, with a corresponding FSR= 541.78(4) MHz. In this setup, light is mode-matched to the fundamental transverse (TEM$_{00}$) mode of the PBC with a $1/e^2$ intensity waist (radius) of 212 µm at the cavity center. A resonant EOM (Qubig) imposes the required frequency sidebands on the laser light. When the EOM is driven with ≈200 mW of rf power at the PBC FSR frequency, a nearly-optimum (for the purpose of eliminating lineshape asymmetry) modulation index $\delta$=1.26 is obtained. We determine this index by measuring the ratio of the 1st-order sideband peak height to that of the carrier, using a separate FP. For the optimum $\delta$=1.26, this peak height ratio is ≈0.64, and sidebands up to 2nd-order are observed in the FP spectrum. All spectral components of the phase-modulated light are injected into the PBC.

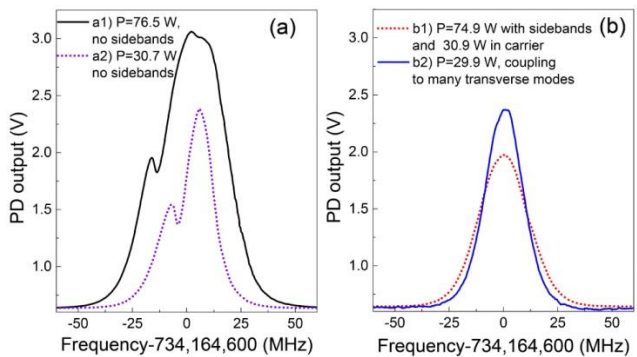

Fig. 3 (Color online). Spectral profiles of the $^{174}$Yb $^1S_0 \rightarrow\ ^3D_1$ transition recorded by scanning the Ti:Sapphire frequency in the following cases: a1) only the carrier is injected into the PBC with 76.5 W in the TEM$_{00}$ mode; a2) same as in a1) but with 30.7 W circulating; b1) with injection of multiple frequencies into TEM$_{00}$ mode and a total of 74.9 W circulating, of which 30.9 W in the carrier ($\delta$=1.26); b2) PBC operation with transverse-mode degeneracy and 29.9 W circulating. The modulation index for b1) was adjusted until the corresponding lineshape was as symmetric as possible. The slight residual asymmetry in b1) is due to the inability to make fine-enough adjustments to the EOM rf driver power-output level, and that of b2) reflects the compromise between the competing requirements for large power coupled to the PBC and minimum residual lineshape distortion.

In the studies employing the degeneracy of the PBC transverse modes, we remove the EOM from the setup, adjust the mode-matching optics for a collimated input beam (≈ 1 mm diameter) to the PBC, and set the PBC mirror spacing to L≈29.3 cm. In this configuration, the various TEM modes are grouped into N=4 fringes and the corresponding FSR is 127.9±0.1 MHz. Although a precise determination of the cavity waist is difficult to make, we do observe a transmitted beam diameter approximately equal to that of the input beam.

Fig. 3 presents the main result of this work. It shows profiles of the $^1S_0 \rightarrow\ ^3D_1$ transition recorded with single-frequency coupling to the TEM$_{00}$ mode, as well as profiles recorded with PBC configurations corresponding to the two methods used to eliminate the profile asymmetry. We observe a nearly complete suppression of the profile distortion in both methods. Comparing the two, PBC operation in multiple longitudinal modes has the relative advantage that the resulting lineshape is long-term stable, since the condition for a non-varying intracavity intensity profile only depends on the modulation index $\delta$. On the contrary, the lineshape when many transverse modes of the FP are excited has a slight dependence on the alignment of the beam coupled to the PBC. On the other hand, the observed 26.8 MHz FWHM linewidth in the first method is larger than that of the second method (20.1 MHz) since although only ≈41% of the circulating power is resonant with the atoms, all of the light contributes to line-broadening due to the ac-Stark effect. As a reference, the linewidths of the asymmetric profiles of cases a1) and a2) in fig. 3, are 38.3 MHz and 20.9 MHz, respectively.

In conclusion, we have demonstrated two methods to eliminate the lineshape distortion of a weak atomic transition in an atomic beam, excited with a strong standing wave field. One method relies on injecting phase-modulated light into the standing wave resonator, and the other on operating the resonator in a configuration in which the degeneracy of transverse modes allows coupling of the input light into many of these modes.

## Acknowledgments

We acknowledge useful discussions with A. Wickenbrock. AMF is supported by the Carl Zeiss Graduate Fellowship.

## References

1. C. E. Wieman, M. C. Noecker, B. P. Masterson, and J. Cooper, Phys. Rev. Lett. **58**, 1738 (1987).
2. J. E. Stalnaker, D. Budker, S. J. Freedman, J. S. Guzman, S. M. Rochester, and V. V. Yashchuk, Phys. Rev. A **73**, 043416 (2006).
3. D. R. Dounas-Frazer, K. Tsigutkin, A. Family, and D. Budker, Phys. Rev. A **82**, 062507 (2010).
4. K. Tsigutkin, D. Dounas-Frazer, A. Family, J. E. Stalnaker, V. V. Yashchuk, and D. Budker, Phys. Rev. Lett. **103**, 071601 (2009).
5. D. Antypas, A. Fabricant, L. Bougas, K. Tsigutkin, and D. Budker, Hyperfine Interact. **238**, 21 (2017).
6. Jun Ye, Long-Sheng Ma, and John L. Hall, J. Opt. Soc. Am. B **15**, 6 (1998).
7. A. Foltynowicz, F.M. Schmidt, W. Ma, and O. Axner, Appl. Phys. B **92**,313 (2008).
8. M. Thorpe, and J. Ye, Appl. Phys. B **91**, 397 (2008).
9. F. Adler, M. J. Thorpe, K. C. Cossel, and J. Ye, Annu. Rev. Anal. Chem. **3**, 175 (2010).
10. S. M. Ball, and R.L. Jones, Chem. Rev., **103**, 5239 (2003).
11. M. J. Thorpe, K.D. Moll, R. J. Jones, B. Safdi, and J. Ye, Science **311**, 1595 (2006).
12. Seung Koo Lee, Jae Jin Kim, and D. Cho, Phys. Rev. A **74**, 063401 (2006).
13. P. Hamilton, M. Jaffe, J.M. Brown, L. Maisenbacher, B. Estey, and H. Müller, Phys. Rev. Lett. **114**, 100405 (2015).

14. S.C. Bennett, Ph. D. thesis, University of Colorado (1996).
15. A. Yariv, and P. Yeh, Photonics, Oxford University Press (2007).
16. D. Budker, S.M. Rochester, and V.V. Yashchuk, Rev. Sci. Instr. **71**, 2984 (2000).
17. D. Budker, D.F. Kimball, and D. P. DeMille, Atomic physics: an exploration through problems and solutions, $2^{nd}$ ed., Oxford University Press (2008).
18. G. C. Bjorklund, Opt. Lett. **5**, 15 (1980).